\documentclass[%
 reprint,
nobibnotes,
 amsmath,
 amssymb,
 aps,
 prl,
]{revtex4-1}

\usepackage{graphicx}% Include figure files
\usepackage{dcolumn}% Align table columns on decimal point
\usepackage{bm}% bold math
\usepackage{gensymb} %has \degree
\usepackage[colorlinks=true,linkcolor=blue,anchorcolor=blue,citecolor=blue,urlcolor=blue]{hyperref}% add hypertext capabilities
\usepackage[mathlines]{lineno}% Enable numbering of text and display math
\begin{document}

\preprint{APS/123-QED}

%\title{Demonstration of extended interaction length in a dielectric laser accelerator using a pulse front tilted laser}
\title{Enhanced energy gain in a dielectric laser accelerator using a tilted pulse front laser}

\author{D. Cesar}
\email{dcesar@ucla.edu}
\affiliation{Department of Physics and Astronomy, UCLA, Los Angeles, California 90095, USA}

\author{J. Maxson}
\altaffiliation[Present address: ]{Department of Physics, Cornell University, Ithaca NY 14850, USA}
\affiliation{Department of Physics and Astronomy, UCLA, Los Angeles, California 90095, USA}

\author{P. Musumeci}
\affiliation{Department of Physics and Astronomy, UCLA, Los Angeles, California 90095, USA}

\author{X. Shen}
\affiliation{Department of Physics and Astronomy, UCLA, Los Angeles, California 90095, USA}

\author{R.\,J. England}
\affiliation{SLAC National Accelerator Laboratory, Menlo Park, California, 94025, USA}

\author{K.\,P. Wootton}
\affiliation{SLAC National Accelerator Laboratory, Menlo Park, California, 94025, USA}

\author{S. Tan}
\affiliation{Department of Applied Physics, Stanford University, Stanford, California, 94305, USA}

\date{\today}% It is always \today, today,

%%%%%%%%%%%%%%%%%%%%%%%%%%%%%%%%%%%%%%%%%%%%%%%%%%%%%%%%%%%%%%%%%%%%%%%%%
\begin{abstract}
Using an 800~nm, 45~fs pulse-front-tilted laser we demonstrate a record 315\,keV energy gain in a dual grating dielectric laser accelerator (DLA) and average accelerating gradients of 560\,MV/m over 0.5\,mm. These results open a new regime in DLA characterized by significant evolution of the beam distribution in the longitudinal phase space, corresponding to $>1/4$ of a synchrotron oscillation. By tilting the laser wavefront we control the resonant velocity of the DLA and observe a net energy gain, indicating that a tapered optical phase could be used to achieve very high energy gain.

\end{abstract}
%%%%%%%%%%%%%%%%%%%%%%%%%%%%%%%%%%%%%%%%%%%%%%%%%%%%%%%%%%%%%%%%%%%%%%%%%
\maketitle
%%%%%%%%%%%%%%%%%%%%%%%%%%%%%%%%%%%%%%%%%%%%%%%%%%%%%%%%%%%%%%%%%%%%%%%%%
%\section{Introduction}
%
% Context ---------------------------------------------------------------
Optical-scale dielectric structures powered by infrared lasers are rapidly being developed into novel low-cost electron accelerators. Driven by compact, high repetition-rate, short-pulse lasers these accelerators could find applications ranging from medical and industrial linacs, to x-ray light sources and particle colliders~\cite{england_dielectric_2014}. At the same time, DLAs intrinsically possess unique attractive features such as attosecond temporal bunch structure, ultrasmall transverse emittances and precise synchronization to external lasers~\cite{wootton_towards_2017}.

DLA, in contrast to many other laser-driven acceleration schemes, is characterized by an accelerating gradient that is proportional to the applied electric field. This is accomplished by using a periodic dielectric structure to diffract an incident plane wave into evanescent fields which form a longitudinal accelerating mode ~\cite{plettner_proposed_2006,plettner_electromagnetic_2011,szczepkowicz_guided-mode_2016, niedermayer_beam_2017}. The structure periodicity $\lambda_g$ is selected to match the phase velocity of space harmonic $n$ to the electron velocity $v$. For a laser incident perpendicular to the electron acceleration axis, the corresponding phase-matching condition is $\lambda_g = \beta n \lambda$, where $\lambda$ is the laser wavelength in vacuum, $\beta = v/c$ and $c$ is the speed of light.

\begin{figure}[htb!]
\centering
\includegraphics[width=2.5 in]{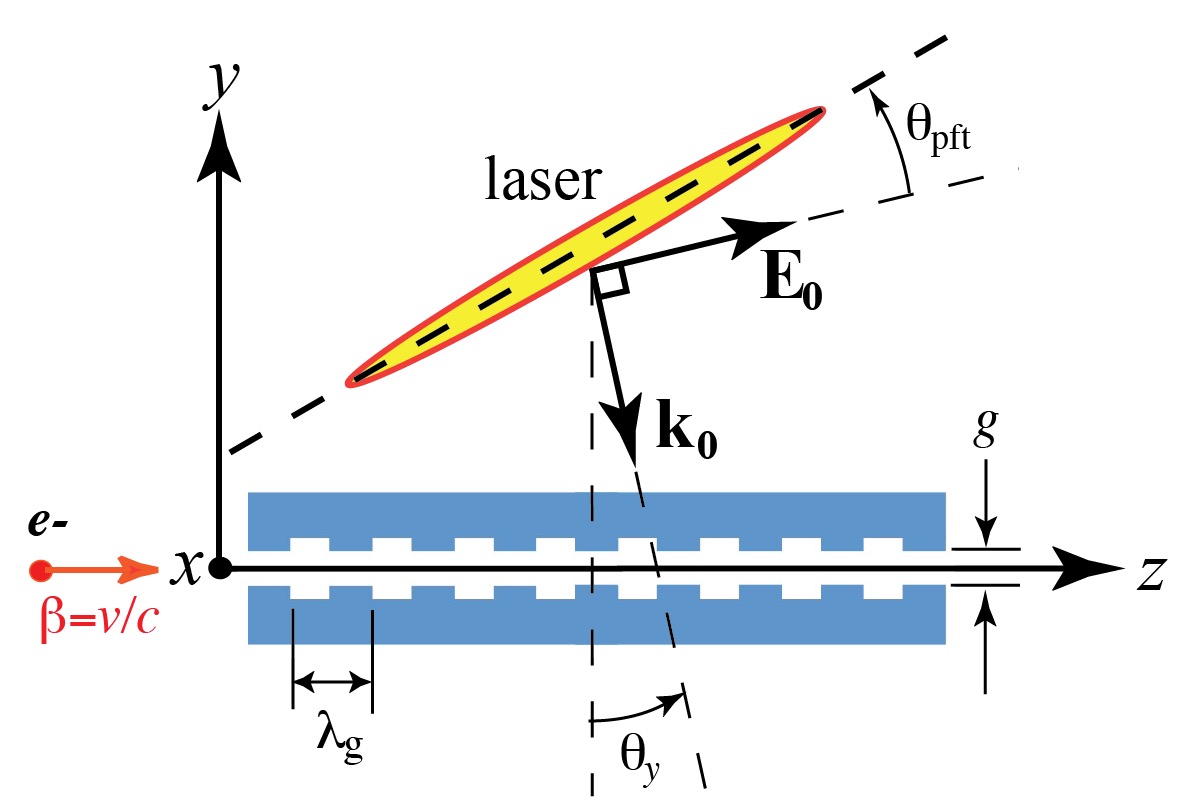}
\caption{Schematic of DLA orientation illuminated by PFT laser, with field amplitude $E_0$ and central wave number $k_0$, incident in the $y$-$z$ plane at angle $\theta_y$. The PFT angle $\theta_\text{pft}$ is group-velocity matched to an electron traveling in the $z$-direction.}
\label{fig:cartoon}
\end{figure}

% Statement of the problem ----------------------------------------------
Efficient coupling to the accelerating mode is possible using a side-coupled laser geometry where the drive laser is polarized along the acceleration axis; but this implies that the laser group velocity is not co-aligned with the electron beam. Thus in all previous DLA experiments ~\cite{peralta_demonstration_2013, breuer_laser-based_2013, leedle_dielectric_2015, wootton_demonstration_2016, kozak_dielectric_2017, kozak_acceleration_2017} the electrons slip out of the laser pulse envelope in approximately one laser pulse duration. And since the peak accelerating gradient is constrained by the material damage threshold, we favor ultrashort laser pulse durations which imply short interaction lengths. Several techniques have been proposed to extend the interaction, including the use of a tilted pulse front~\cite{plettner_proposed_2006,wei_dual-grating_2017}, multi-stage guided wave coupling~\cite{hughes_-chip_2017,koyama_parameter_2014}, and powering the structure by a co-propagating laser~\cite{naranjo_stable_2012}.

In the present work we demonstrate electron acceleration using a DLA with an interaction length 50 times longer than the laser pulse duration by using a pulse-front tilted (PFT) illumination scheme.This is significant because the resulting interaction length is not only long in terms of the number of optical periods, but also in terms of the longitudinal phase advance of the accelerating electrons. At the observed gradients, an interaction length of 0.5\,mm corresponds to $>1/4$ synchrotron oscillation in the accelerating wave potential, enabling for the first time the study of longitudinal dynamics of relativistic electrons in a DLA. Taking advantage of the PFT-enabled long interaction length, we show that tuning the resonant energy by tilting the laser wavefront yields a net energy gain for an initially unbunched beam.

PFT works in a DLA by matching the laser envelope to the particle trajectories, so energy gain can occur along the entire length of the accelerator. With reference to Fig.~\ref{fig:cartoon}, a coupling is introduced between $z$ and $t$ so that the laser peak intensity illuminating the vacuum channel ($z$ axis) can be described as $I(z,t) \propto f\left(z/c-\beta_g t\right)$ where $f(t)$ represents the temporal profile of the un-tilted laser pulse and $\beta_g = \cos \theta_\text{pft} \csc (\theta_\text{pft}+\theta_y)$ is the effective normalized group velocity. If $\beta_g c$ is matched to the electron velocity, the pulse envelope appears stationary in the electron beam frame. For speed-of-light electrons with normal incidence laser ($\theta_y = 0$), the desired $\theta_\text{pft} = 45^\circ$.

\begin{figure}[htb!]
\centering
\includegraphics[width=\columnwidth]{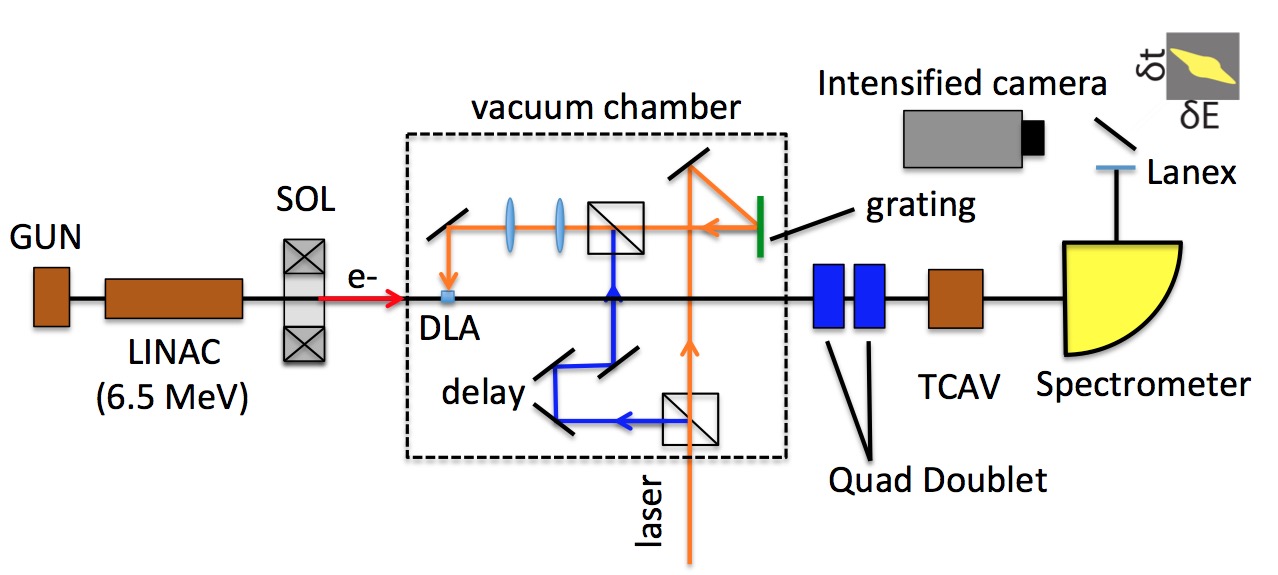}
\caption{Schematic (not to scale) of experimental setup. A 6.5~MeV energy electron beam from the UCLA Pegasus gun-linac \cite{alesini_new_2015} is focused into the DLA by a solenoid (SOL). After PFT acceleration in the DLA, the electron beam is recollimated by a quadrupole doublet, deflected in $y$ by a deflecting cavity (TCAV) and then in $x$ by a dipole (Spectrometer), permitting the energy-time phase space of the beam to be imaged with an intensified camera (upper right inset).}
\label{fig:experiment}
\end{figure}

The optical system used to create the PFT pulse for the present experiment is detailed in Ref.~\cite{cesar_optical_2018}, and is based on similar techniques developed for single-cycle THz generation by optical rectification~\cite{hebling_generation_2008}. A 600\,line/mm diffraction grating (see Fig. \ref{fig:experiment}) introduces PFT which is tuned by imaging the grating onto the DLA with a variable magnification telescope. Since the DLA is very sensitive to the drive laser phase, the PFT optics must be engineered to not only image the grating plane~\cite{kostenbauder_ray-pulse_1990}, but also to ensure a flat phase-front to minimize phase slippage of the electrons. The PFT angle and pulse duration are measured at the interaction $x$-$z$ plane by cross-correlating the PFT laser with a non-tilted replica of the drive laser sent through a variable delay line (indicated in blue in Fig.~\ref{fig:experiment}) while adjusting the variable magnification telescope. To account for the 6.5\,MeV electron energy $(\beta_g < 1)$, $\theta_\text{pft} = 45.3 \pm 0.5^\circ$ was used in this experiment.

Three DLA structures (Table~\ref{parametertable}) were used in the experiment. The structures were fabricated by etching pillars (height 700\,nm, width 325\,nm) into 500\,$\mu$m thickness fused silica wafers~\cite{peralta_design_2012}. A structure with smaller vacuum gap $g$ results in a higher accelerating gradient $G$ (for the same incident electric field $E_0$), while shorter structures are less sensitive to dephasing.

\begin{table}[htb!]
\centering
\caption{Geometries and results for structures A, B, and C.}
\begin{tabular}{lccccc}
  \hline
    Parameter 		& Symbol 		& Units  	& A 	& B 	& C \\
  \hline
Structure length 	    & $L$ 			&  mm 		& 0.5  	&  1.0  & 0.5  \\
Vacuum gap 		    & $g$ 			&  nm 		& 800  	& 800   & 400  \\
Structure factor    & $\kappa$      &           & 0.08  & 0.08  & 0.11 \\
\hline
Incident electric field	& $E_0$ 	    &  GV/m     & 7     & 7     & 7 \\
Interaction length    	& $L_\text{int}$&  mm 		& 0.47 	&  0.7 	& 0.47 \\
Max energy gain 	    & $\Delta E_\text{max}$ 	&  keV 		&  250  & 315   & 280  \\
Average gradient 	    & $G$			&  MeV/m  	&  500  & 315   & 560 \\
  \hline
\end{tabular}
\label{parametertable}
\end{table}

\begin{figure}[htb!]
\centering
\includegraphics[width=\columnwidth]{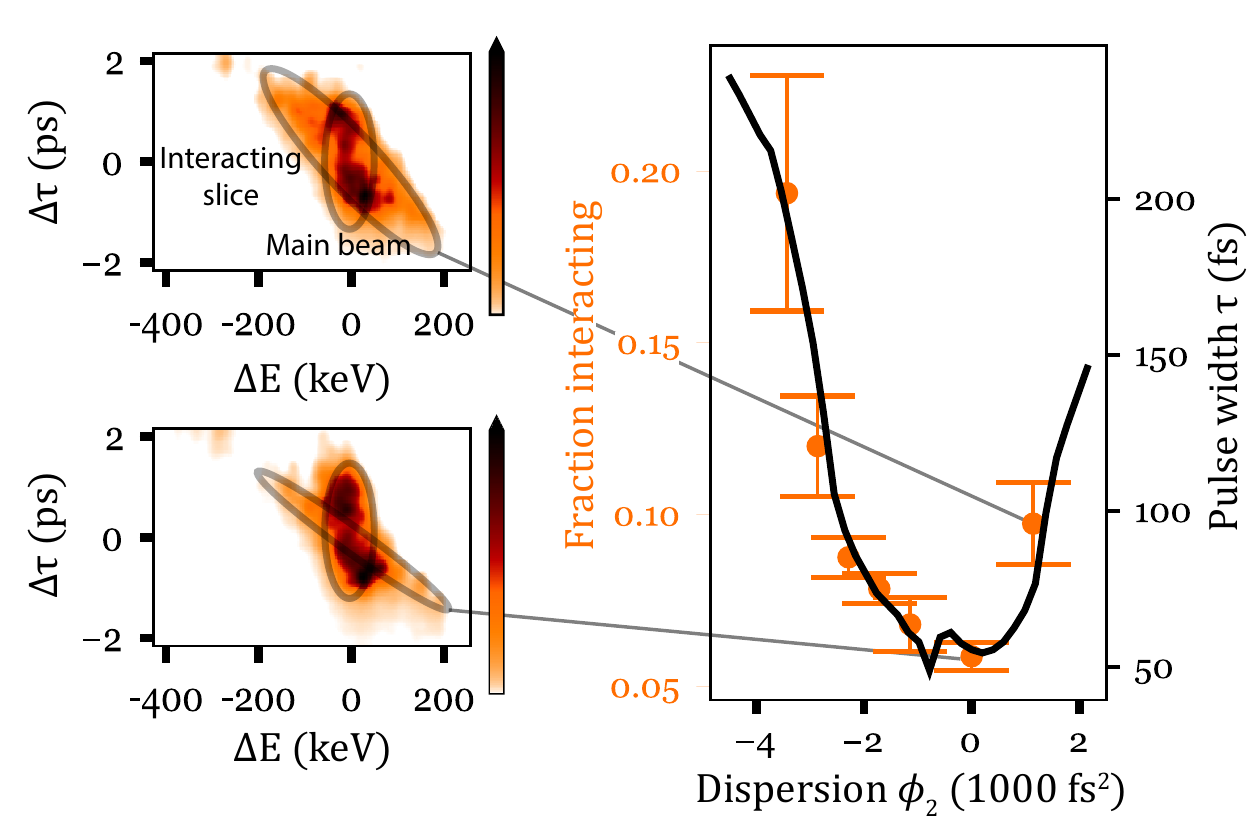}
\caption{Measured phase space images (a, b) show that the DLA accelerates electrons in a temporal ``slice" of the main beam.  The fraction of interacting electrons is shown in (c) to track the measured laser pulse duration $\tau$ (in orange) as the laser compressor dispersion $\phi_2$ is adjusted.}

\label{fig:fractionaccelerated}
\end{figure}

The vertical angular aperture $\Delta y' = g/L$ of the vacuum channel is maximized for Structure A, but even in this case the focused electron beam (with root-mean-square (RMS) spot size and angular spread $\sigma_y = 10\,\mu$m, $\sigma_{y'}$ = 1\,mrad) overfills the aperture.  Consequently, 2\% of the particles are transmitted, and the rest strike the bulk wafer, scatter, and lose energy (median 290\,keV, mean 400\,keV). Some of the scattered particles are collected by the spectrometer and are visible in the top left of the spectra in Fig.~\ref{fig:fractionaccelerated}.

Among the transmitted electrons only a short temporal slice interacts with the PFT laser since the electron bunch duration (1\,ps RMS) is significantly longer than the full width half-max (FWHM) laser pulse duration ($\tau = 45$\,fs). To observe this effect, we recorded images of the time-dependent modulation of the beam energy by using a transverse deflecting cavity (``TCAV" in Fig.~\ref{fig:experiment}) to streak along $y$ followed by a magnetic spectrometer to disperse the energy distribution along $x$. The resulting $x$-$y$ transverse profile on the screen represents the beam longitudinal phase space, as shown in Fig.~\ref{fig:fractionaccelerated}(a,b) where the DLA interaction appears as a temporal slice of accelerated electrons.  The slice is correlated in time and energy due to time-of-flight difference during the 1.5\,m drift from the DLA to the TCAV. Without PFT the images would show no such correlation, since the entire electron bunch would see the same acceleration.

%%%%%%%%%%%%%%%%%%%%%%%%%%%%%%%%%%%%%%%%%%%%%%%%%%%%%%%%%%%%%%%%%%%%%%%%%

We increase the fraction of electrons which interact with the laser (at the cost of laser intensity) by adjusting a grating compressor to add dispersion to the input laser pulse.  In Fig.~\ref{fig:fractionaccelerated}(c) we show that this fraction (measured with the TCAV off as the fraction of the total charge for which $|\Delta E|>\Delta E_\text{max}/\sqrt{2}$) closely tracks the laser pulse duration (solid curve) measured via frequency-resolved optical gating.

\begin{figure}[htb!]
\centering
\includegraphics[width=\columnwidth]{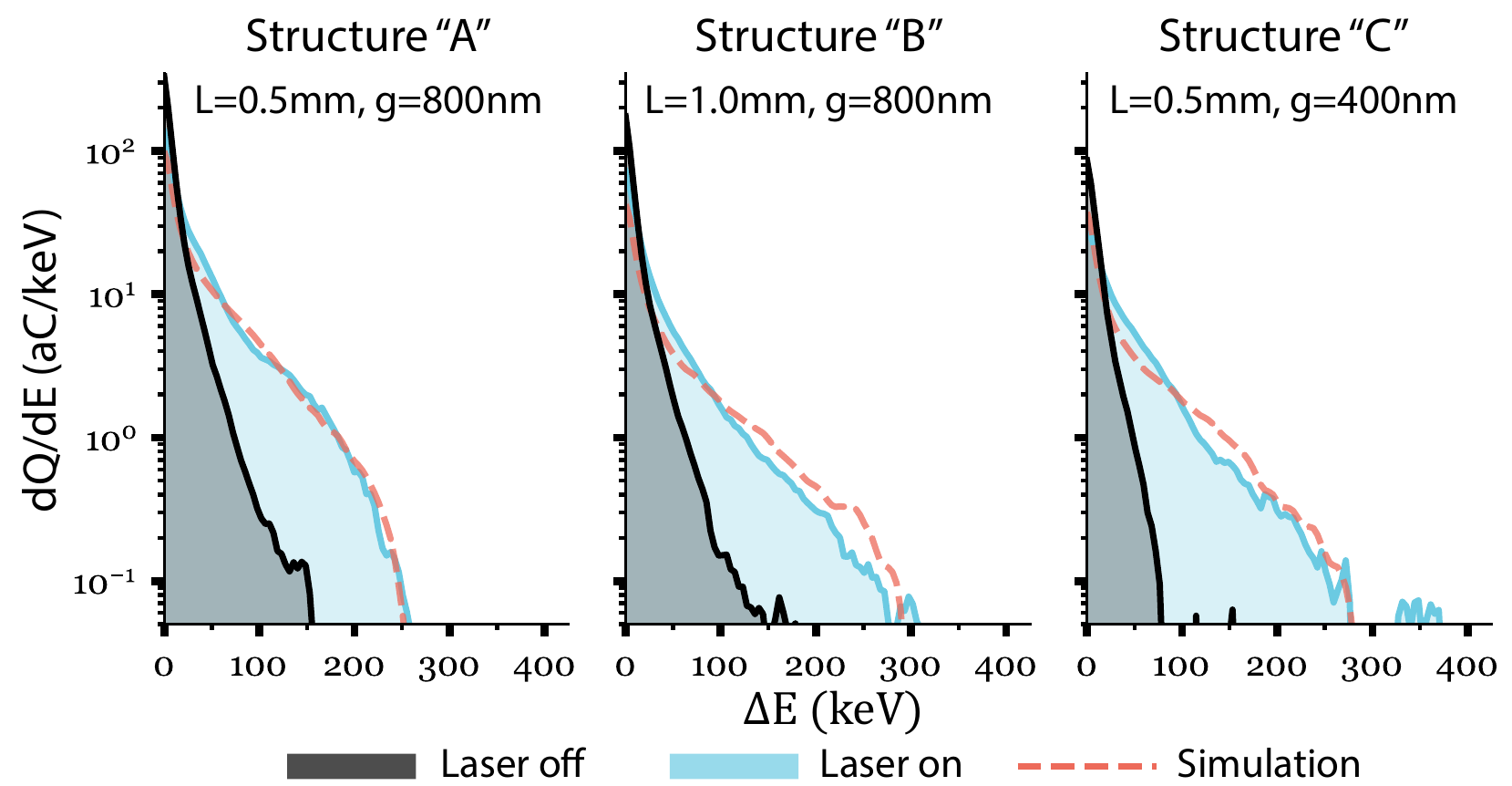}
\caption{Measured electron energy spectra with the laser off (black) and laser on (light blue) for the three different accelerators of Table~\ref{parametertable}.  Particle tracking simulations are also shown (red dashed). Each spectrum is an average of 50 shots.}

\label{fig:spectra}
\end{figure}

To better resolve the acceleration produced by each of the three structures we measure the electron beam energy spectrum  with the TCAV turned off, as shown in Fig.~\ref{fig:spectra}. The most prominent feature is the main peak of non-interacting electrons, followed by a small fraction of accelerated electrons reaching out to a cutoff energy of $\Delta E_\text{max} \approx \kappa E_0 L_\text{int}$, where $\kappa$ is a property of the DLA representing the ratio of peak gradient $G_p$ to incident field ($\kappa \equiv G_p/E_0$) and $L_\text{int}$ is the interaction length.

Simulated spectra (dashed curves in Fig. \ref{fig:spectra}) are calculated by tracking particles through the field of an initially Gaussian transverse laser distribution which has been propagated through the PFT optics and a nonlinear model of the DLA substrate~\cite{cesar_nonlinear_2017}. The structure factor $\kappa$ for each gap was calculated using an electromagnetic solver~\cite{noauthor_fdtd_nodate}. 

Observed energy gains and gradients are summarized for the three structures in Table~\ref{parametertable}. The highest energy gain of 315\,keV (Structure B) has the lowest average gradient since the interaction length is shortened from the full structure length $L$ = 1\,mm to $L_\text{int} = 0.7$\,mm by a laser spot size $w_z = 0.65$\,mm ($1/e^2$ in intensity) and a small misalignment in $z$ of the laser axis relative to the center of the DLA. The Gaussian spatial distribution of laser intensity further lowers the average gradient due to dephasing from the nonlinear Kerr effect. The results of Table \ref{parametertable} are consistent with the expectation that average gradient is increased when $L < w_z$ and/or if a narrower gap $g$ is used.  Consequently, the highest gradient ($G$ = 560\,MeV/m) is observed in Structure C, which satisfies both conditions.

The PFT scheme allows us to advance from accelerating a 1\,ps electron bunch for $\sim20$\,$\mu$m~, as in prior work \cite{cesar_nonlinear_2017, wootton_demonstration_2016}, to accelerating a 50\,fs bunch for 0.5\,mm. Hence, we greatly increase the interaction length while simultaneously utilizing the high gradients enabled by short laser pulse durations. The observed maximum energy gain is more than an order of magnitude greater than previous demonstrations using fs laser pulses without PFT~\cite{wootton_demonstration_2016, cesar_nonlinear_2017}.

With an increased interaction length the electrons undergo dynamic evolution in the accelerating wave, which we demonstrate by tuning the resonant phase velocity in the structure. The phase of an electron in a Floquet harmonic of the DLA is $\Psi=n k_g z-\omega t+\phi(z)$, where the phase $\phi(z) = (\omega/c) \sin\left(\theta_y\right) z$ is a linear phase shift caused by tilting the wavefront of the drive laser. Thus, the normalized phase velocity of the DLA is $\beta_\text{res} = \omega / k_\text{res} c $, where $k_\text{res} = (\omega/c) \sin \theta_y + k_g$ (using $n=1$ for the resonant harmonic).

Using $\Psi$, the electron energy gain can be heuristically approximated by
\begin{equation}
\Delta E \approx \kappa E_0 \int_0^L \sin \left(\Psi(z,t(z))\right)dz,
\label{eqn:kinematicintegral}
\end{equation}
and further simplified by assuming a constant electron velocity such that $t(z) = (z-z_0)/\beta c t$ is the arrival time of a particle at position $z$. Maximum energy gain is achieved when $\Psi$ is constant, i.e. electron beam velocity $\beta$ is matched to the structure phase velocity $\beta_{res}=\omega/\left[k_g+(\omega/c) \theta_y\right]$.
Viewed in the frequency domain, a PFT thus allows the electron to interact with a broadband pulse (centered around $\omega_0$) by providing the angular dispersion $\theta_y(\omega)=\theta_0+\tan(\theta_\text{pft})(\omega_0/\omega)$ required to make all frequency components travel at the same velocity. We can then vary the offset $\theta_0$ by tilting the DLA structure to tune the phase velocity of the accelerating mode and control the dephasing length.

\begin{figure}[htb!]
\centering
\includegraphics[width=\columnwidth]{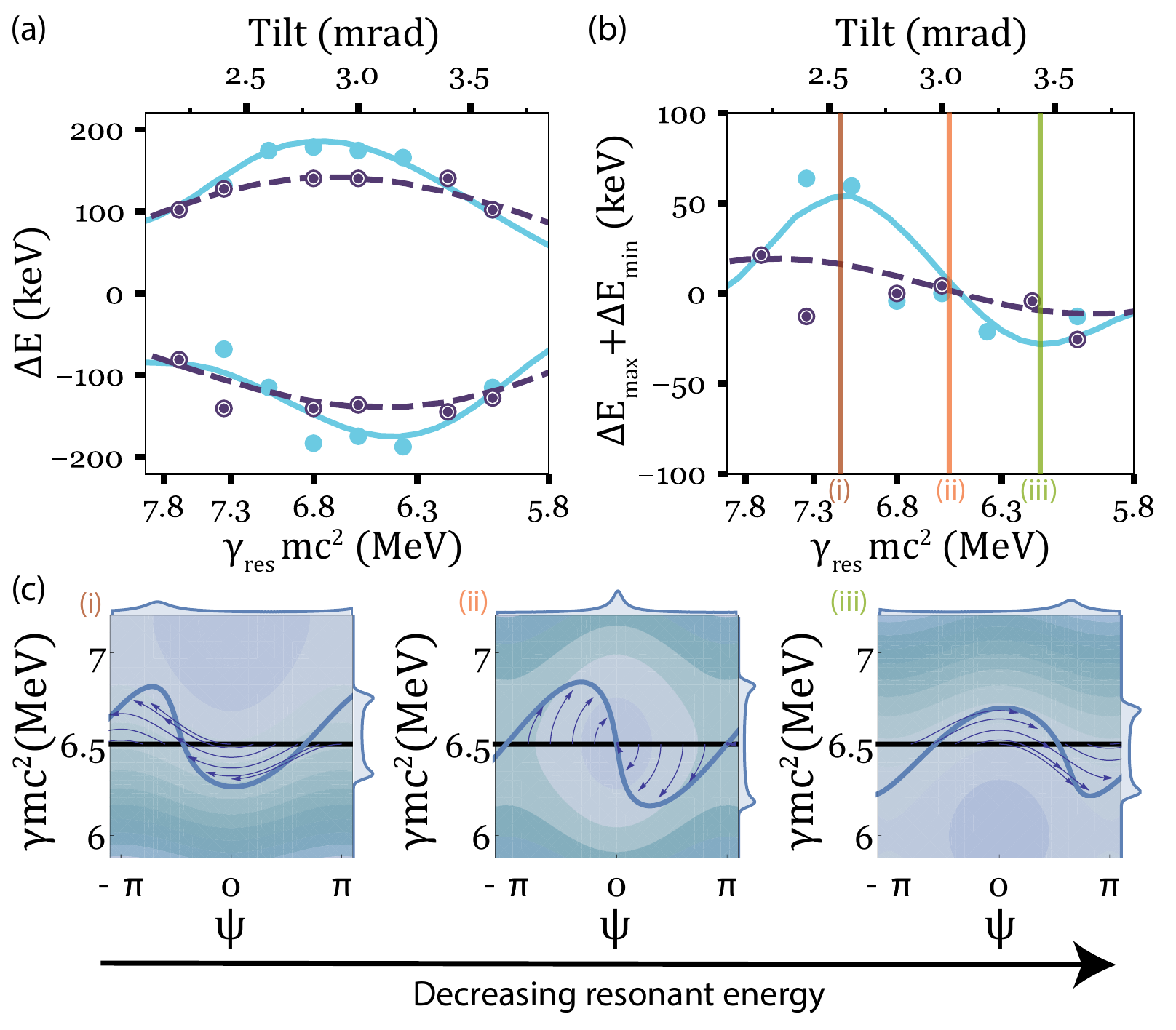}
\caption{Longitudinal dynamics in a DLA.  Maximum energy gain and energy loss (at top and bottom of the plot respectively) are shown in (a) and their difference in (b) as a function of $\theta_y$ for structures A (purple) and B (blue). Dots are measurements and lines are simulation. Computed Hamiltonian dynamics are shown in (c) for structure B, corresponding to the three resonant energies marked by vertical lines in part (b).  Each plot shows the potential energy (contours), the initial (black) and final (blue) beam distributions, individual trajectories (arrows), and histograms of the final distribution (top and right)}
\label{fig:tiltscan}
\end{figure}

The effect of dephasing on the measured maximum and minimum energy changes ($\Delta E_\text{max}$, $\Delta E_\text{min}$) was measured as a function of $\theta_y$ for the two $g=800$\,nm structures ``A'' and ``B'', as shown in Fig.~\ref{fig:tiltscan}. The ideal tilt for our structure is offset from zero because the structure was designed for speed-of-light electrons ($\lambda_g = \lambda = 800\,\textrm{nm}$, and $\beta_{res} = 1$ at $\theta_y = 0$). For the present experiment ($\lambda = 803$\,nm, $\beta$ = 0.9969), the optimal $\theta_y = 3.1$\,mrad. When varying $\theta_y$ around this value we observe that the particle-wave resonance is sharper in ``B'' ($L=1$\,mm) than in ``A'' ($L=0.5$\,mm) because in the longer channel the electrons have more periods over which to dephase. Quantitatively, the energy gain can be estimated by integrating Eq.~(\ref{eqn:kinematicintegral}) to obtain $\Delta E \propto \text{sinc}⁡(\omega/c \theta_0 L)$ but this only adequately describes the energy gain in the shorter $L=0.5$\,mm accelerator.

For the $L=1$\,mm accelerator there is an asymmetric energy gain/loss which can be interpreted by moving beyond the kinematic approximation to consider a one-dimensional normalized relativistic Hamiltonian expanded around the resonant energy $\gamma_\text{res} \equiv (1-\beta_\text{res}^2)^{-1/2}$:
\begin{equation}
H(\eta,\Psi)\approx \frac{\eta^2}{2\gamma_{res}^2}+\frac{\alpha}{\gamma_{res}} \sin⁡(\Psi)
\label{eqn:hamiltonian}
\end{equation}
where $\eta \equiv (\gamma-\gamma_\text{res})/\gamma_{\text{res}}$ and $\alpha \equiv e G /\left(k_\text{res} m c^2\right)$ is similar in magnitude to a radiofrequency (rf) proton linac \cite{wangler_standard_2008} or small-signal free-electron laser~\cite{lee_accelerator_2011}.

Phase space trajectories for this Hamiltonian are shown in Fig.~\ref{fig:tiltscan}(c) for three different resonant energies. The largest energy modulation is observed near $\gamma_\text{res} m c^2 = 6.5$\,MeV, but for other injection energies, the energy gain and loss are asymmetric due to the beam phase advance. In particular, when the beam is injected below the resonant velocity it gains energy, whereas when the beam is injected above the resonant velocity it loses energy. This is the origin of the asymmetry in the data that is seen by comparing the difference between the maximum energy gain and the maximum energy loss [Fig.~\ref{fig:tiltscan}(b)].

An estimate for the tilt angle yielding maximum net energy gain can be obtained by analyzing the evolution of a nearly-monochromatic unbunched beam in the Hamiltonian (Eq. ~\ref{eqn:hamiltonian}). This system is analogous to a small signal free electron laser and the spectrum will be shifted up or down according to the small signal gain function $F(\zeta)=\frac{1}{4}\frac{\partial}{\partial \zeta}\text{sinc}^2(\zeta/2)$, where $\zeta \equiv \eta k_\text{res} z / \gamma_\text{res}^2$~\cite{lee_accelerator_2011}. Maximum energy gain occurs when $\eta_\text{max}= 0.41 \gamma_\text{res}^2/N \approx 0.12$, in our case where $N = L_\text{int}/\lambda_g = 875$. In terms of resonant energy $mc^2 \gamma_{res,m} = mc^2 \gamma_\text{res}(1 + \eta_\text{max})$ = 7.3\,MeV in excellent agreement with the data in Fig.~\ref{fig:tiltscan}(b). This approach neglects a small correction due to the Kerr-induced nonlinear phase modulation ~\cite{cesar_nonlinear_2017} which is included in the simulation lines shown in Fig.~\ref{fig:tiltscan}(a,b) and is responsible for the larger net energy gain than net energy loss.

The impact of the drive laser wavefront on DLA longitudinal dynamics suggests that a ``soft tapering'' of the interaction can be imposed by shaping the transverse phase profile of the laser in the PFT setup. For example, imaging a programmable liquid crystal phase mask onto the DLA would allow nearly arbitrary phase profiles to be applied, resulting in full control of dynamics in the DLA. Compared to physically varying the periodicity of the structure \cite{lemery_adiabatic_2017, hanuka_trapping_2017}, the ``soft-tapering'' offers a tunable solution for future DLAs.

In conclusion, we present the first demonstration of a DLA powered by a PFT laser pulse. Increasing the interaction length by an order of magnitude resulted in a record energy gain of 315\,keV. The resonant velocity of the accelerating mode was tuned by tilting the laser illumination angle, resulting in spectra with net electron energy gain or loss. For longer DLA structures, these results suggest that imprinting phase profiles upon the laser pulse could be used to control the longitudinal phase space evolution of the beam, enabling MeV-scale energy gains in a single acceleration stage.

%%%%%%%%%%%%%%%%%%%%%%%%%%%%%%%%%%%%%%%%%%%%%%%%%%%%%%%%%%%%%%%%%%%%%%%%%
%\section{Acknowledgement}
\begin{acknowledgments}
This work was supported in part by the Gordon and Betty Moore Foundation under grant GBMF4744 (Accelerator on a Chip), and in part by National Science Foundation under grants PHY-1734215 and PHY-1535711 and U.S. Department of Energy, Office of Science, under Contracts no. DE-AC02-76SF00515 and DE-SC0009914.
\end{acknowledgments}
%%%%%%%%%%%%%%%%%%%%%%%%%%%%%%%%%%%%%%%%%%%%%%%%%%%%%%%%%%%%%%%%%%%%%%%%%
% Bibliography

\bibliographystyle{apsrev4-1} % Tell bibtex which bibliography style to use
\bibliography{pftpaperbib}

\end{document}